\documentclass[12pt]{article}
\usepackage{amsmath, amssymb, fullpage, epsfig}
\usepackage{graphicx}
\usepackage{shadow} 
\usepackage[latin1]{inputenc}
\usepackage{dsfont}
\usepackage{color}
\usepackage{multicol}
\usepackage{enumitem}
\usepackage{tikz}
\usetikzlibrary{arrows,snakes,shapes.arrows,decorations.markings}
     \tikzset{>=triangle 90}
     \tikzstyle{Bl}=[draw,circle,blue,scale=.6]
     \tikzstyle{bl}=[draw,circle,blue,fill=blue,scale=.35]
     \tikzstyle{R}=[draw,circle,fill=red,scale=.6]
     \tikzstyle{bbc}=[draw,circle,fill=black,scale=.75]
\usepackage{hyperref}
\usepackage{float}
\usepackage{fancyhdr}
\usepackage{tensor}
\usepackage{slashed}
\hypersetup{colorlinks=true}

\usetikzlibrary{calc,trees}

\newcommand\be{\begin{equation*}}
\newcommand\ee{\end{equation*}}
\newcommand\bea{\begin{eqnarray*}}
\newcommand\eea{\end{eqnarray*}}
\newcommand\del{\partial}

\def\bar{\overline}

\def\d{{\delta}}

\definecolor{dark-green}{rgb}{0,0.5,0}
\definecolor{mywhite}{rgb}{1,1,1}
\definecolor{myblack}{rgb}{0,0,0}
\definecolor{darkgray}{gray}{0.5}
\definecolor{lightgray}{gray}{0.75}

\newcommand\pubnumber{DPF2015-351}
\newcommand\pubdate{\today}

\def\cincy{Department of Physics\\
University of Cincinnati, Cincinnati, OH45220}
\def\support{\footnote{Work supported by Department of Energy under grant DE-SC0011784.}}

\def\Title#1{\begin{center} {\Large #1 } \end{center}}
\def\Author#1{\begin{center}{ \sc #1} \end{center}}
\def\Address#1{\begin{center}{ \it #1} \end{center}}

\newcommand\pubblock{\rightline{\begin{tabular}{l} \pubnumber\\
         \pubdate  \end{tabular}}}
\newenvironment{Abstract}{\begin{quotation}  }{\end{quotation}}
\newenvironment{Presented}{\begin{quotation} \begin{center} 
             PRESENTED AT\end{center}\bigskip 
      \begin{center}\begin{large}}{\end{large}\end{center} \end{quotation}}
\def\Acknowledgments{\bigskip  \bigskip \begin{center} \begin{large}
             \bf ACKNOWLEDGMENTS \end{large}\end{center}}

\def\beq{\begin{equation}}
\def\eeq#1{\label{#1}\end{equation}}
\def\eeqn{\end{equation}}

\def\beqa{\begin{eqnarray}}
\def\eeqa#1{\label{#1}\end{eqnarray}}
\def\eeqan{\end{eqnarray}}

\let\bar=\overbar

\def\Dslash{\not{\hbox{\kern-4pt $D$}}}
\def\dslash{\not{\hbox{\kern-2pt $\del$}}}

\def\ee{e^+e^-}

\def\msb{{\bar{\ssstyle M \kern -1pt S}}}

\begin{document}
\begin{titlepage}
\pubblock

\vfill
\Title{On the classification of planar rank-1 d=4 N=2 superconformal field theories}
\vfill
\Author{ Yongchao L\"u\support}
\Address{\cincy}
\vfill
\begin{Abstract}
We describe an approach to classifying four-dimensional conformal field theories
with N=2 supersymmetry and a Coulomb branch of vacua with the topology of the
complex plane\cite{allm15}.  We also discuss the Higgs/mixed branches and conformal/flavor central charges.
\end{Abstract}
\vfill
\begin{Presented}
DPF 2015\\
The Meeting of the American Physical Society\\
Division of Particles and Fields\\
Ann Arbor, Michigan, August 4--8, 2015\\
\end{Presented}
\vfill
\end{titlepage}
\def\thefootnote{\fnsymbol{footnote}}
\setcounter{footnote}{0}

\section{Introduction}

The idea of treating QFTs as universal mathematical objects is motivated by the application of QFTs in describing diverse physical systems.  Then the following natural questions arise:\\
\indent $\bullet$ How do we precisely define and characterize them?\\
\indent $\bullet$ How can we classify them?\\
\indent In very broad terms, QFTs can be organized into different subclasses by their space-time dimension, and the presence of supersymmetries (SUSYs) and internal global symmetries.  In this report, we will be interested in d=4 QFTs with N=2 superconformal symmetry, or d=4 N=2 superconformal field theories (SCFTs). It is well known that most of the d=4 N=2 SCFTs admits no conventional Lagrangian description\cite{Argyres:1995jj,Argyres:1995xn,mn9608}. However,  it turns out that the N=2 superconformal symmetry leads to significant constraints on the structures of the theories, like the operator algebras and correlation functions.  Especially, the N=2 SCFTs can be characterized by a relatively small set of data.  Hopefully, this data can be used to \emph{uniquely} define a theory.

However, this data is notably difficult to solve for purely on grounds of symmetry considerations.  So it is hard to pursue a general classification which reflects the intrinsic definition of N=2 SCFT. 

In fact there is one merit of N=2 superconformal symmetry which sheds light on the classification problem.  It is that the N=2 superconformal symmetry also enables powerful theoretical control over the dynamics of the system and leads to exact results for various interesting physical observables.  Quite often those exact observables encode important and tractable information about the SCFT data.

One important example is the so called Seiberg-Witten (SW) data.  By the Seiberg-Witten data we just mean the collection of low-energy observables which can be exactly predicted by N=2 supersymmetry\cite{sw9408}.  As we will review below, this is encoded in a geometrical object, called the Coulomb branch (CB). We call the dimension of the CB the ``rank" of the CB.  Various N=2 SCFT data enter the geometry of the Coulomb branch in rather non-trivial ways. This motivates us to attack the N=2 classification problem by directly considering the classification of Seiberg-Witten geometries of the Coulomb branches.  By further imposing certain physical constraints on the low energy effective dynamics, we will show in section 2 the result of the classification of a subset of N=2 SCFTs, namely, the planar rank-1 d=4 N=2 SCFTs.

Once we have established the existence of these theories, it is desirable to understand other interesting physical observables.  In section 3, we will discuss two subjects, namely, Higgs/mixed branches and conformal/flavor central charges.  They can often be exactly determined by various available methods.

\section{Seiberg-Witten geometry and classification of \\d=4 N=2 SCFTs}

Let's firstly review certain general features of d=4 N=2 SCFTs and N=2 RG flows.  Any N=2 SCFT has a global $U(1)_R \times SU(2)_R$ symmetry, and one can organize the operators into different protected sectors according to their $R$-charge assignments.  For our purposes, we will be interested in the following four types of operator supermultiplets: Coulomb branch operators, Higgs branch operators, energy-momentum tensor operators, and conserved flavor current operators.  The reason for this choice lies in the fact that the vevs of Coulomb branch operator and Higgs branch operators parametrize the moduli space of N=2 vacua, and the operator product coefficients of the energy-momentum tensor/conserved flavor current operators give the conformal/flavor central charges. 

All the possible deformations of N=2 SCFTs by N=2 preserving local couplings span the N=2 theory space. Note that the local deformations in the vicinity of an N=2 fixed point can be classified as irrelevant, exactly marginal, or relevant deformations.  While we will be mainly interested in the exactly marginal and relevant cases for the sake of their connection with IR dynamics, there could be subtle issue related to  ``dangerously" irrelevant operators which develop into extra relevant operators in the IR fixed point theory.  However, we have argued in \cite{allm15} that such a possibility is absent for the N=2 RG flow, and then we need only consider the so-called ``safely" irrelevant deformations.  Also, we will mainly discuss the mass deformation associated with moment map operator ($\in \mathfrak{f}^*$) of the flavor current, and their associated mass parameters $m_i$ which have scaling dimension $D(m_i)=1$. Upon flavor symmetry transformation, mass parameters are valued in the Cartan subalgebra of the complexified flavor algebra $\mathfrak{f}_{\mathbb{C}}$.

The moduli space of N=2 vacua consists of several branches, each factorized into a ``Coulomb branch factor" and a ``Higgs branch factor". This local product structure is implied by the N=2 selection rules on the low energy dynamics\cite{argyressusynote}.  We will call the branch without Higgs branch factor as the Coulomb branch, and the complex dimension of Coulomb branch as the rank.  Now let's consider the N=2 SCFTs with a 1-complex-dimensional Coulomb branch homeomorphic to the complex plane; we will refer to them as planar ``rank-1" d=4 N=2 SCFTs.  Then the low energy dynamics in Coulomb branch is the subject of Seiberg-Witten theory, as described below.

The spontaneously broken scale symmetry (and related $U(1)_R$ symmetry) implies a $\mathbb{C}^*$ action on the Coulomb branch which gives the rank-1 case the structure of a complex cone with defect angle.  The Coulomb branch vev $u$ has positive scaling dimension $D(u)$.   Unbroken N=2 supersymmetry, electric-magnetic duality, and the Dirac quantization condition imply a rigid special K\"ahler (RSK) geometry on the CB which can be described in terms of a fibration of complex tori over the CB.  Furthermore, the singularity associated with degenerate tori lies at the origin of Coulomb branch and can be determined by the conjugacy class of monodromies in the EM duality group taken by the loop surrounding the singularity.  The full list of the possible singularities is given by Kodaira's $ADE$ classification of singular elliptic fibrations over the complex plane\cite{kodaira}.\footnote{Note that the terminology $ADE$ is related to the Dynkin diagram associated with the intersection form of the blow-up cycles.} 

Upon deforming the SCFT by generic mass terms, the scaling symmetry and flavor symmetries are both broken explicitly.  The Coulomb branch is not a complex cone anymore, and the singularity will be split into a group of undeformable singularities. The deformation by non-generic mass terms will also beinteresting to us, for they can connect different N=2 SCFTs through RG flow by taking a proper scaling limit around the particular singularity.

A powerful algebraic recipe associated with Seiberg-Witten theory is the so-called Seiberg-Witten curve and (meromorphic) one-form\cite{sw9408}.  The curve contains the Weyl-invariant homogeneous polynomials of the mass parameters.  The one-form is required to satisfy rigid special K\"ahler (RSK) conditions constraining its $u$- and $m$-dependence.  By a specific ansatz the one-form encodes the flavor symmetry by summing over the several Weyl orbits on the weight spaces\cite{mn9608,nty9903}. Then the further consideration of scaling limit and RG flows at various non-generic mass patterns, together with the ``safely" irrelevant deformation conjecture, will result in a subset of the planar Seiberg-Witten geometries as the consistent low energy theories of N=2 SCFTs\cite{allm15}.

In the end, we finish the classification of planar rank-1 N=2 SCFTs, and they can be uniformly labeled by their CB scaling dimension $D(u)$ and flavor symmetry algebra $\mathfrak{f}$, as $\bold{\text{SCFT}}[D[u], \mathfrak{f}]$. Among them, there are two interesting series depending on their flavor symmetry pattern:
\begin{itemize}
\item $\mathfrak{f}$ is $ADE$ type, that is the $ADE$-series.  The Coulomb branches of those theories correspond to the maximal deformation of the Kodaira singularities. All of them admit N=2 S-duality \cite{as0711,aw0712} and in turn ``class-S" constructions \cite{gaiotto0904,cdt1212}. They also have F-theory constructions\cite{Sen:1996vd,Banks:1996nj,dhiz9812}.

\item $\mathfrak{f}$ contains an $\mathfrak{sp}$ factor; we call these theories the ``$\mathfrak{sp}$-series". The Coulomb branches of these theories correspond to the submaximal deformation of the Kodaira singularities. All of them also admit N=2 S-duality and in turn class-S construction, while no known F-theory construction exists.

\end{itemize}








\section{Higgs/mixed branches and conformal/flavor \\central charges}

Higgs/mixed branch structures and conformal/flavor central charges are important N=2 SCFT data. Both of them can be related to the N=2 OPE algebra, as described below.  In the case of theories admitting S-dual descriptions, one can obtain exact results about them.  Furthermore, there exists a general method applicable to the theories without S-dual description for computing central charges by considering the topological twisted Coulomb branch partition function and Seiberg-Witten data\cite{Shapere:2008zf}.

\paragraph{Higgs/mixed branch.}

The ADE-series of rank-1 theories admit F-theoretic constructions in terms of world volume theories on one D3 brane parallel to a 7-brane system of type ADE and therefore their Higgs branches can be interpreted as the moduli spaces of centered single instantons of ADE gauge groups\cite{Gaiotto:2008nz}.  In turn it is well known that the centered one-instanton moduli space of group $F$ is equivalent to the minimal nilpotent orbit of group $F_{\mathbb{C}}$\cite{Kronheimer:1990ay}.  The minimal nilpotent orbit description makes explicit the complex symplectic structure of the Higgs branch.  In particular, the Joseph ideal --- which is the quadratic polynomial defining minimal nilpotent orbit as subvariety in $F_{\mathbb{C}}^{*}$ --- can actually be understood as the OPE relation for the Higgs branch chiral ring.  

One may hope to generalize this connection and derive the Higgs/mixed branch structure from N=2 chiral ring relations directly.  However, the flavor representations for the Higgs branch operators involved for mixed branches are not adjoint representation generally, so we need go beyond coadjoint orbits and consider more general cases involving the orbit structure of flavor group action on general representations.  Again, the N=2 chiral ring relations can be used to derive quadratic polynomial defining relations for the Higgs branch factor of the mixed branch.  By further considering the OPE relation with Coulomb branch operator, one may deduce the fibration structure of the mixed branch\cite{allm15}. 

Now let's turn to the theories admitting S-dual descriptions.  They involve the conformal manifolds --- i.e., the spaces of exactly marginal couplings --- of N=2 superconformal lagrangian theories, and the isolated rank-1 SCFTs emerge when the complexified gauge couplings are taken to specific limits.  It is often possible that various coupling-independent properties of isolated strongly-coupled SCFTs can be deduced from the weakly-coupled lagrangian description.  As mentioned above, ADE-series and $\mathfrak{sp}$-series SCFTs admit S-duality constructions, and therefore their Higgs/mixed branches can be exactly determined.  Due to the space limitations, we will redirect the interested reader to our work \cite{allm15} for more details.

\paragraph{Conformal/flavor central charges.}

We now turn to the computation of the conformal/flavor central charges of rank-1 N=2 SCFTs.  For the theories admitting S-dual descriptions, while the conformal central charges can be determined readily, the flavor central charge can only be determined for the semi-simple parts of the flavor symmetry due to the lack of suitable normalization for the Dynkin index for the abelian flavor symmetry factors.  

There exists another method by Shapere and Tachikawa for computing these central charges from the low energy data on the CB\cite{Shapere:2008zf}.  It also requires the flavor symmetry group to be semi-simple.  In it, an N=2 SCFT is put on a curved manifold with a non-trivial metric and a background gauge field for the flavor symmetry group $F$ via the twisting of the $SU(2)_{R}$ with one of the $SU(2)$'s of $SU(2) \times SU(2)\simeq SO(4)$ of the tangent bundle.  A topologically twisted sector of the theory is still protected by a supersymmetry such that one can relate the various central charges of the UV theories to various IR data.   The dimension of the mixed branch and the action of the flavor group on it turn out to be important inputs into this computation\cite{allm15}.

\section{Summary}

In conclusion, we have shown the existence and classified a set of d=4 N=2 SCFTs with planar Coulomb branches, and obtained interesting information about their Higgs/mixed branches and conformal/flavor central charges.  Here are several future directions:
\begin{itemize}
\item Certain puzzles about the computation of the flavor central charges for the abelian symmetry are still remain, and it would be interesting to develop new methods to address them.

\item The F theory construction for the $ADE$-series can give the flavor symmetry manifestly by a string junction/web construction\cite{dhiz9812}. It would be interesting to obtain similar constructions for other rank-1 theories, especially for the $\mathfrak{sp}$-series.

\item It will be interesting to extend our methods to explore the classification of N=2 SCFTs of higher ranks, i.e., by considering Coulomb branch singularities and RSK deformations. 
\end{itemize}

\Acknowledgments
It is a pleasure to thank P. Argyres, M. Lotito, and M. Martone for collaboration on the project\cite{allm15}.  I am especially grateful to P. Argyres for helpful comments on this manuscript.  


\end{document}